\documentstyle[sprocl,epsfig]{article}

\def\be{\begin{equation}}
\def\ee{\end{equation}}
\def\bea{\begin{eqnarray}}
\def\eea{\end{eqnarray}}
\def\xp{x_{{I\!\!P}}}

\def\zmin{z_{\mbox{\scriptsize{min}}}}
\def\pt{p_{T}}
\def\ptt{p^{2}_{T}}
\def\kt{k_{T}}

\def\kt{k_T}

\def\xprime{x^{\prime}}

\def\xprim2bar{\overline{x}^{\prime\prime}}
\def\ptbold{\mbox{\boldmath$p$}_T}
\def\ktbold{\mbox{\boldmath$k$}_T}

\def\gapprox{\lower .7ex\hbox{$\;\stackrel{\textstyle >}{\sim}\;$}}
\def\lapprox{\lower .7ex\hbox{$\;\stackrel{\textstyle <}{\sim}\;$}}

\begin{document}

\title{DIFFRACTIVE DIJET PHOTOPRODUCTION AND\\
       THE OFF-DIAGONAL GLUON 
       DISTRIBUTION\footnote{Presented by K.~Golec--Biernat
at the 6th International Workshop on Deep Inelastic Scattering and
 QCD (DIS98), Brussels, 4-8 April 1998.}}

\author{K. GOLEC--BIERNAT$^{1,2}$, J.KWIECI\'NSKI$^{2}$,  
           A.D.MARTIN$^{1}$}

\address{$^{1}$Department of Physics,
University of Durham, Durham DH1 3LE, England\\
$^{2}$H. Niewodnicza\'nski Institute of Nuclear Physics, Krak\'ow, Poland}

%%%%%%%%%%%%%%%%%%%%%%%%%%%%%%%%%%%%%%%%%%%%%%%%%%%%%%%%%%%%%%
% You may repeat \author \address as often as necessary      %
%%%%%%%%%%%%%%%%%%%%%%%%%%%%%%%%%%%%%%%%%%%%%%%%%%%%%%%%%%%%%%

\maketitle\abstracts{Diffractive dijet photoproduction is proposed as a
probe of the off-diagonal gluon distribution and its 
evolution.
Predictions for the transverse momentum distribution of the jets are given.
Differences with DGLAP evolution are highlighted.}

Off-diagonal 
({\it non-forward~\cite{rad}, off-forward~\cite{ji}},
{\it non-diagonal~\cite{cfs}}) 
parton distributions (OFPD's)
are generalizations of
the conventional (diagonal) parton distributions. 
While the latter are related
to the diagonal matrix elements of  the twist-two quark or gluon operators 
$\langle p | \ldots | p \rangle$, 
the OFPD's characterize 
the  matrix elements $\langle p' | \ldots | p \rangle$ 
between the nucleon states with different momenta~\cite{rad,ji}.
We consider the unpolarized case and
suppress  polarization among the nucleon  state
characteristics. 
Although the OFPD's in general cannot be regarded
as particle densities, they
provide important information about the 
nonperturbative structure of the nucleon~\cite{ji}. 
They are also indispensable in description of such processes as
deeply virtual Compton scattering~\cite{rad,ji} and hard diffractive
electroproduction of vector mesons~\cite{rad,cfs,vecmes,mr}. 
These processes  offer
a possibility to determine the essentially nonperturbative OFPD's. 
We propose another process, exclusive 
diffractive photoproduction of dijets with
high values of transverse momenta, as a particularly good probe of the
OFPD's. The necessity for renormalization of the 
quark and gluon operators in the definition of the OFPD's leads to
evolution equations just  as for the conventional parton distributions. 
The process we propose allows the study of off-diagonal evolution 
in a  kinematical range 
which is not probed by the two other processes. 
Thus our process significantly increases
the possibility of the experimental determination of the OFPD's.

\section{CROSS SECTION OF THE PROCESS}

The amplitude for our process is shown in Fig.~1.
Two jets with high values of transverse momenta $\pm \mbox{\boldmath $p$}_T$
are produced through the exchange of  two gluons 
with  longitudinal momentum fractions $x$ and $\xprime$ and transverse
momentum $\ktbold$. 
%With the indicated flow of the gluon momenta
The first gluon is emitted from the proton, and the second is absorbed
if $\xprime >0$. However
the  situation with $\xprime <0$ is not forbidden; 
in this case
the second gluon is emitted rather than absorbed~\cite{rad,ji}.
%and it is convenient to reverse its momentum flow in Fig.~1. 
In both cases
the $t$-channel momentum transfer $r=(x-\xprime)~p \equiv \xp~p$, and
$t=r^2=0$. 

%%%%%%%%%%%%%%%%%%%%%%%%%%%%%%%%%%%%%%%%%%%%%%%%%%%
%FIGURE 0
\begin{figure}[t]
   \vspace*{-1cm}
    \centerline{
     \epsfig{figure=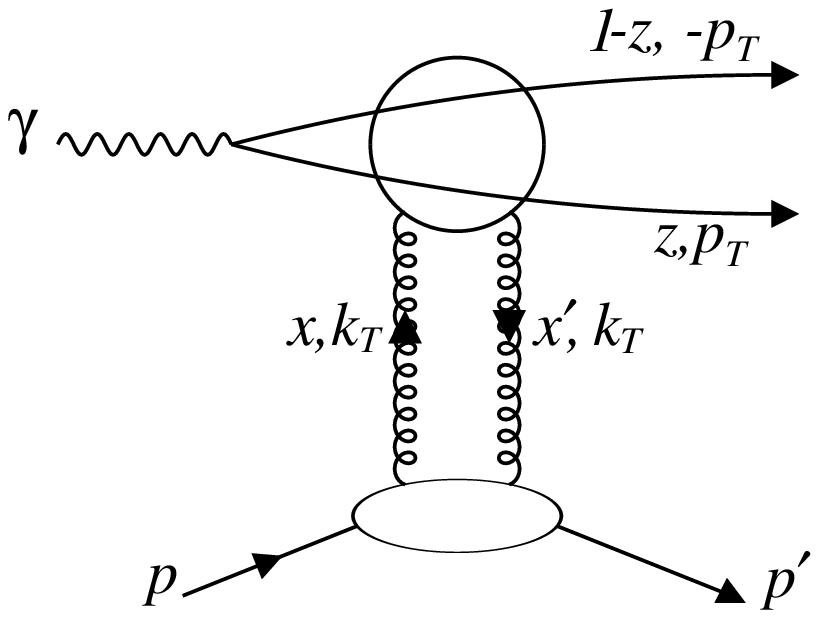,height=5.5cm}
               }
    \vspace*{-0.5cm}
     \caption{The amplitude for the exclusive
     diffractive dijet photoproduction.}
\label{fig:0}
\end{figure}
%%%%%%%%%%%%%%%%%%%%%%%%%%%%%%%%%%%%%%%%%%%%%%%%%%%

The cross section for the process is obtained by squaring the 
amplitude approximated by its imaginary part, taking into account 
the four different ways that the two gluons 
can couple to the quarks/jets. The quarks
are assumed to be massless. Thus we obtain \cite{nz,gkm}
\be
\label{eq:1}
\left.
 \frac{d \sigma_{T}}{d^2 \ptbold  dt}
\right|_{t = 0} 
 =  \frac{\alpha \alpha_S^2}{6 \pi \ptt} \;
  \sum_q e_q^2
\int\limits_{\zmin}^{1-\zmin} dz\;  [z^2 + (1 - z)^2 ]\; 
[\phi_1(z,\pt)+\phi_1(1-z,\pt)]^2\;,
\ee
where $z$ is the fraction of the photon momentum  carried by the quark. 
The impact factor $\phi_1$ is given by
\be
\label{eq:2}
\phi_1(z,\pt) = \frac{\pi}{\pt^2}\;\int \frac{d \tau}{\tau^3}\;
\int_0^{\pi} \frac{d \phi}{\pi}\; 
\frac{\partial G(x,\xprime,k_T^2)}{\partial \ln k_T^2}\;
\frac{1}{2}
\biggl\{1-\frac{1-\tau^2}
{1+\tau^2 + 2 \tau \cos \phi}
\biggr\}\;,
\ee
where $\tau=\kt/\pt$ and  $G(x,\xprime,k_T^2)$ is the off-diagonal 
gluon distribution 
in which the  longitudinal momentum fractions are given by
\be
\label{eq:4}
x=\xp + \xprime\;,~~~~~~~~~~~
\xprime=\frac{\pt^2}{z W^2}\;\bigl\{\tau^2+2 \tau \cos \phi \bigr\}\;.
\ee
Here $\phi$ is the angle between the transverse momentum vectors
$\ktbold$ and $\ptbold$, and $\xp=M^2/W^2$ 
with the diffractive mass of the dijet
system $M^2=\pt^2/(z(1-z))$. The fraction 
$x$ is always positive, and for $\tau < 2$ it varies in the range
\be
\label{eq:5}
z\;\xp\;<\;x\;<\;1\;,
\ee
which leads to both  $\xprime>0$ (for $x>\xp$) and $\xprime<0$ (for $x<\xp$).
Thus  we study the OFPD's in the full kinematical
range of the $x$ and $\xprime$ variables.

\section{EVOLUTION EQUATIONS AND DIJET CROSS SECTION}

%%%%%%%%%%%%%%%%%%%%%%%%%%%%%%%%%%%%%%%%%%%%%%%%%%%%%%%%%%%%%%%%%%%%%%
%FIGURE 4
\begin{figure}[t]
   \vspace*{-1cm}
    \centerline{
     \epsfig{figure=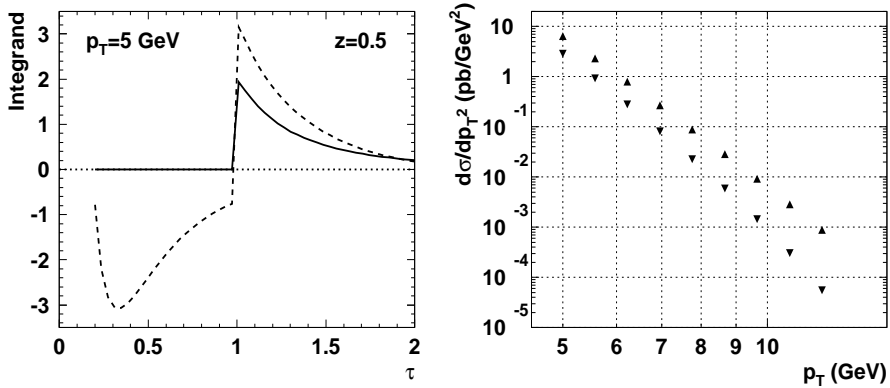,height=7cm,width=13.5cm}
               }
    \vspace*{-0.5cm}
     \caption{The integrand of (\ref{eq:2}),  and cross section (\ref{eq:1})
      integrated over $t$ with $e^{6 t}$, for different analyses.
      The solid curve and the lower points correspond to
      the simplified diagonal case and the 
      dashed curve and the upper points to the full off-diagonal
      analysis.
     }
\label{fig:4}
\end{figure}
%%%%%%%%%%%%%%%%%%%%%%%%%%%%%%%%%%%%%%%%%%%%%%%%%%%%%%%%%%%%%%%%%%%%%%

\begin{figure}[t]
   \vspace*{-1cm}
    \centerline{
     \epsfig{figure=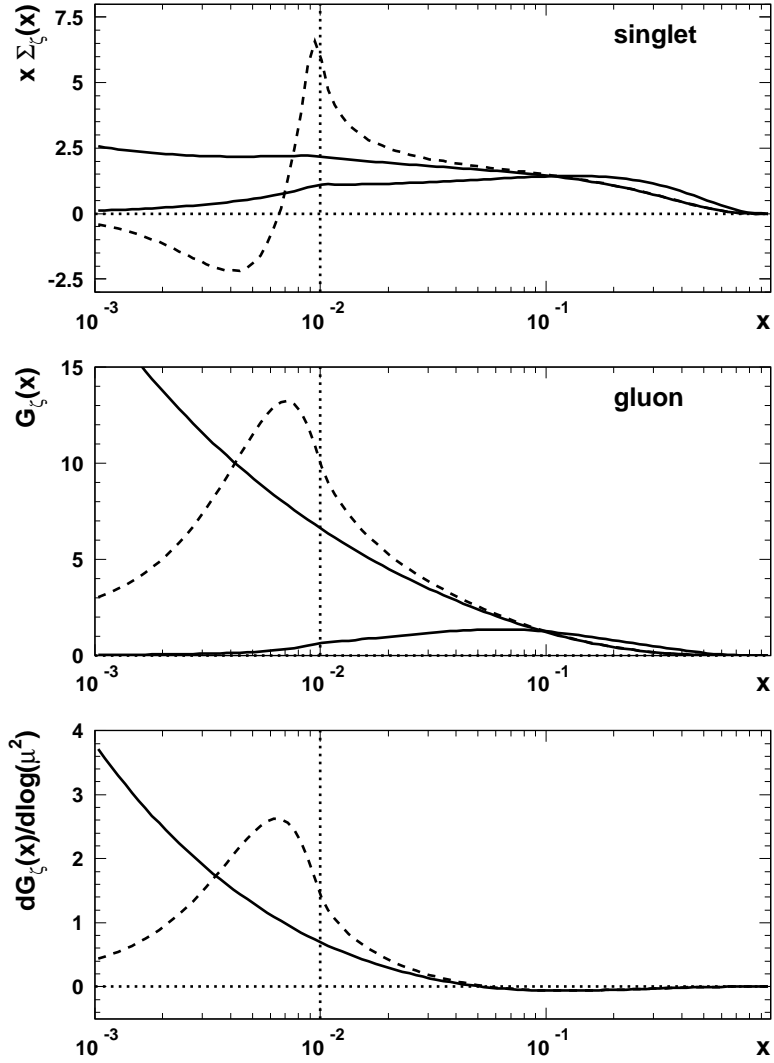,height=15cm,width=13cm}
               }
    \vspace*{-0.5cm}
     \caption{The OFDP's at $\mu^2=10^2~\mbox{\rm GeV}^2$  for 
     $\zeta=\xp=10^{-2}$ (dashed curves). The solid lines
     show the initial distributions at $\mu_0^2=1~\mbox{\rm GeV}^2$
     (lower curves) and effect of their DGLAP evolution to
     the same value of $\mu^2$ (upper curves).     
}
\label{fig:3}
\end{figure}

The OFPD evolution equations do not mix different values
of $\zeta\equiv x-\xprime=\xp$. Thus it is convenient
to change the notation to $G_{\zeta}(x) \equiv G(x,\xprime)$ \cite{rad}.
After introducing the collective notation 
${\cal{F}}_{\zeta}(x)\equiv(\Sigma_{\zeta}(x),\;G_{\zeta}(x))$ 
for the singlet and gluon off-diagonal distributions
the evolution equations~\cite{rad,ji,mr,evol} have the form
\be
\label{eq:6}
\mu \frac{\partial}{\partial \mu}\;{\cal{F}}_{\zeta}(x,\mu) =
\int_0^1 dz\; {\cal{P}}_{\zeta}(x,z;\mu)\;{\cal{F}}_{\zeta}(z,\mu)\;.
\ee
Their explicit form can be found in \cite{gkm}. 
Eqs.~(\ref{eq:6}) combine features of the DGLAP evolution 
equations~\cite{dglap} 
for $x>\zeta$ and the ERBL evolution equations~\cite{erbl} for partonic
distribution amplitudes for $x<\zeta$. 
This is shown in Fig.~{\ref{fig:3}} where the off-diagonal evolution for
the singlet $\Sigma_{\zeta}$, 
gluon $G_{\zeta}$ and $\partial G_{\zeta}/\partial \log(\mu^2)$ distributions
(dashed curves) is compared with the DGLAP evolution (upper solid curves).
The chosen initial distributions (lower solid curves) reflect the mixed 
nature of the off-diagonal distributions
\be
\label{eq:7}
{\cal{F}}_{\zeta}(x,\mu_0)\;=\;\theta(x-\zeta)\; {\cal{F}}_{AP}(x,\mu_0)
\;+ \;\theta(\zeta-x)\; {\cal{F}}_{BL}(x,\mu_0)\;
\ee
and ensure the necessary condition  ${\cal{F}}_{\zeta}(0)=0$~\cite{rad}.
The recent MRST pa\-ra\-me\-tri\-za\-tion~\cite{mrst} is used for 
${\cal{F}}_{AP}$, and ${\cal{F}}_{BL} \sim x^n(1-x)^n$ where $n>0$.

In Fig.~{\ref{fig:4}} we show the integrand of (\ref{eq:2}), 
and the cross section (\ref{eq:1}), 
for different assumptions about the gluon distribution and its evolution.
In the first simplified case, shown by the solid lines,  the diagonal
$G_{\zeta=0}(\xp)$ distribution is
evolved with the DGLAP equations from input (\ref{eq:7}).
The angular integration in (\ref{eq:2}) gives  the theta function 
$\Theta(\tau-1)$ and cross section (\ref{eq:1}) has no contribution from
$\kt<\pt$~\cite{nz}. 
In the second case the off-diagonal gluon distribution $G_{\zeta}(x)$ is
evolved with (\ref{eq:6}) from input (\ref{eq:7}) (dashed lines). 
The main difference lies in
the $\tau<1$ region which now gives an important contribution 
to the cross section.
The effect on the
dijet cross section is shown by comparison of the 
lower points (simplified case)
with the upper ones (off-diagonal analysis). 
Clearly the impact of the true off-diagonal
analysis is significant. The details of the relation between 
the form of the integrand
of (\ref{eq:2}) and the particular behaviour of the OFPD's 
are discussed in~\cite{gkm}.

\section*{Acknowledgments}
Important discussions with  A.V. Radyushkin, M.G. Ryskin and M. W\"{u}sthoff, 
the Royal Society/NATO Fellowship and KBN grant
no. 2 P03B 089 13 are gratefully acknowledged.

%%%%%%%%%%%%%%%%%%%%%%%%%%%%%%%%%%%%%%%%%%%%%%%%%%%%%%%%%%%%%%

\newpage
%%%%%%%%%%%%%%%%%%%%%%%%%%%%%%%%%%%%%%%%%%%%%%%%%%%%%%%%%%%%%%%%%%%%%%
%FIGURE 3
%%%%%%%%%%%%%%%%%%%%%%%%%%%%%%%%%%%%%%%%%%%%%%%%%%%%%%%%%%%%%%%%%%%%%% 

\end{document}